\documentclass[11pt,preprint,superscriptaddress,aps,showkeys,nofootinbib]{revtex4}

\usepackage{amssymb,amsmath,amsfonts}
\usepackage{graphicx}
\usepackage{graphics}
\usepackage{eepic,epsfig}
\usepackage{indentfirst}
\usepackage[utf8]{inputenc}
\usepackage[T1]{fontenc}
\usepackage{subfig}

1.60cm \makeatletter \@addtoreset{equation}{section}

\makeatother

\begin{document}

\title{Quasinormal modes of a massive scalar field nonminimally coupled to gravity in the spacetime of Self-Dual Black Hole} 

\author{J. S. Santos}
\email{sjaianedasilvasantos@yahoo.com}
\affiliation{Departamento de F\'{\i}sica, Universidade Federal de Campina Grande \\
Caixa Postal 10071, 58429-900 Campina Grande, Para\'{\i}ba, Brazil}
\author{M. B. Cruz}
\email{messiasdebritocruz@gmail.com}
\affiliation{Departamento de F\'{\i}sica, Universidade Federal de Campina Grande \\
Caixa Postal 10071, 58429-900 Campina Grande, Para\'{\i}ba, Brazil}
\author{F. A. Brito}
\email{fabrito@df.ufcg.edu.br}
\affiliation{Departamento de F\'{\i}sica, Universidade Federal de Campina Grande \\
Caixa Postal 10071, 58429-900 Campina Grande, Para\'{\i}ba, Brazil}
\affiliation{Departamento de F\'{\i}sica, Universidade Federal da Para\'{\i}ba \\ Caixa Postal 5008, 58051-970, Jo\~ao Pessoa, Para\'{\i}ba, Brazil}
 
\begin{abstract}
In this work, we investigate the quasinormal modes for a massive scalar field with a nonminimal coupling with gravity in the spacetime of a loop quantum black hole, known as the Self-Dual Black Hole. In this way, we have calculated the characteristic frequencies using the 3rd order WKB approach, where we can verify a strong dependence with the mass of scalar field, the parameter of nonminimal coupling with gravity, and parameters of the Loop Quantum Gravity. From our results, we can check that the Self-Dual Black Hole is stable under the scalar perturbations when assuming small values for the parameters. Also, such results tell us that the quasinormal modes assume different values for the cases where the mass of field is null and the nonminimal coupling assumes $\xi=0$ and $\xi=1/6$, i.e., a possible breaking of the conformal invariance can be seen in the context of loop quantum black holes.
\end{abstract}


\maketitle

\section{Introduction}

General Relativity (GR) is accepted as the best description of gravitation physics. One of the most striking predictions of Einstein's theory is the prediction of the black holes (BHs), which are objects from which nothing (even light signals) can escape after crossing the event horizon. The interest in BHs goes beyond astrophysics because they have been pointed out as possible objects that can help us to understand one of the most intriguing problems in theoretical physics nowadays: the conciliation of quantum physics and gravitation through a quantum gravity theory (QG). Because it is expected that in the presence of a very strong gravitational field, the quantum nature of spacetime becomes revealed.

One of the main candidates for a theory of quantum gravity is the Loop Quantum Gravity (LQG) \cite{Rovelli:2004tv}, which competes with string theory \cite{Zwiebach:2004tj}. In the context of LQG, it is possible to get interesting theoretical models that provide an insight into the quantum characteristics of spacetimes revealed by BHs. An important scenario corresponds to the quantum version of the Schwarzschild black hole (SchBH), which is called self-dual black hole (SDBH) \cite{Modesto:2009ve}. The SDBH solution has a very interesting property, self-duality. Wherefrom such property, the physical singularity of BHs can be replaced by an asymptotically flat region, which is an expected effect in a regime of QG.

In the last years, it has increased the interest in BHs physics because of gravitational waves (GWs) observations, originated from a binary BHs merging and neutron stars. This class of events has been observed with great precision through the LIGO and Virgo collaborations \cite{LIGOScientific:2018mvr}. These GWs are due to the perturbations of BHs, once we cannot find BHs completely isolated in the Universe. Thus, the BHs are always in perturbed states due to the interaction with other compact objects in their neighborhood, for instance, other BHs or neutron stars. Also, the perturbations are characterized by a set of complex eigenvalues (frequencies) of the wave equations, called quasinormal modes (QNMs), and, therefore, they can be observed through experiments with gravitational interferometers.

As aforementioned the QNMs are complex values, where the real parts give the oscillation frequencies, while the imaginary parts determine the damping rates. The QNMs of BHs depend only on BHs parameters and not on how they were perturbed. Thus, the QNMs are known, as the “fingerprint” of a BH. The studies of the QNMs are of great interest and importance in different contexts \cite{Horowitz:1999jd, Berti:2009kk}.

In recent years, it has been suggested that the QNMs can play a very important role in understanding quantum aspects of gravity theories. Especially, in the LQG context, it has been also suggested that the QNMs can be used to fix the value of the Immirzi parameter, a parameter measuring the quantum of the spacetime \cite{Dreyer:2002vy}, a fundamental issue that remains open in this field. In this context, the QNMs have been analyzed for the SDBH solution in \cite{Santos:2015gja}, where the authors have considered the perturbations of the massless scalar field. Also, considering gravitational perturbations the analyses have been made in \cite{Cruz:2015bcj, Cruz:2020emz}, where the results showed for the first time in the context of BHs in LQG, a breaking of the isospectrality. Still, in the context of LQG, it is made an analysis of the QNMs for a rotating black hole solution in \cite{Liu:2020ola}.

The present work has the main goal to perform an analysis of the QNMs spectrum for the metric of SDBH. Thus, we will consider the perturbations through a massive scalar field coupled nonminimally with gravity \cite{Wald:1995yp, Birrell:1979ip}. Some efforts in that sense were considered \cite{Konoplya:2002wt, Ohashi:2004wr, Konoplya:2018qov, Gwak:2019ttv, Yekta:2019por}. For instance, in \cite{Konoplya:2018qov} it was analyzed the long-lived quasinormal modes and the instability in a Reissner-Nordström spacetime. The motivations for considering the nonminimal coupling with gravity can be found in different contexts: cosmology \cite{Spokoiny:1984bd, Barvinsky:2008ia, Setare:2008pc, Uzan:1999ch, Kamenshchik:1995ib}, general relativity \cite{Donoghue:1994dn, Allen:1983dg, Ishikawa:1983kz}, superstring theory \cite{Maeda:1985bq}, induced gravity \cite{Accetta:1985du} and loop quantum gravity \cite{Ashtekar:2003zx}. To compute the QNMs we will use the WKB (Wentzel - Kramers - Brillouin) approach, which has been introduced by Schutz and Will \cite{Schutz:1985km}, and after improved by Iyer and Will \cite{Iyer:1986np} and more recently by Konoplya \cite{Konoplya:2003ii, Konoplya:2019hlu}, Matyjasek and Opala \cite{Matyjasek:2017psv} and Hatsuda \cite{Hatsuda:2019eoj}. In this context, we will study the influence of the parameters of LQG, the mass of the scalar field, and the nonminimal coupling with gravity in the QNMs, and consequently in the stability of the SDBH solution.

The paper is organized as follows. In Section \ref{intr_sel_dual_bh}, we briefly review the SDBH solution and discuss their self-duality property. In Section \ref{effect_potent}, we derive a Schrödinger-like equation, where is considered a massive scalar field nonminimally coupled with gravity. In Section \ref{qnm_sdbh}, we calculate the QNMs through the WKB method, and finally, we summarize our results and draw concluding remarks in Section \ref{concluding}. Throughout this work, we use natural units $\hbar=c=G=1$ and metric signature $(-,+,+,+)$.

\section{Self-Dual Black Hole}
\label{intr_sel_dual_bh}

In this section, we will briefly introduce the SDBH solution that arises from a simplified model of LQG. This solution is consisting of an asymmetry-reduced model corresponding to homogeneous spacetimes, and for a more detailed study of the model, see Ref. \cite{Modesto:2009ve}.

The structure of SDBH corresponds to a quantum version of the SchBH and is described by the metric
\begin{equation} \label{lqgbh_metric}
 \begin{aligned}
  ds^2 &= - \frac{(r-r_{+})(r-r_{-})(r+r_{*})^2}{r^{4}+a_{0}^{2}} dt^2 + \frac{dr^2}{\frac{(r-r_{+})(r-r_{-})r^{4}}{(r+r_{*})^{2}(r^{4}+a_{0}^{2})}} + \left(r^{2} + \frac{a_{0}^{2}}{r^{2}} \right) \left ( d\theta^2 + \text{sin}^2 \theta d\phi^2\right ) .
 \end{aligned}
\end{equation}
In the Eq. \eqref{lqgbh_metric}, we have the presence of an external horizon localized in $r_{+}=2m$, an intermediate in $r_{*}=\sqrt{r_{+}r_{-}}$ and a Cauchy horizon localized at $r_{-}=2mP^2$. Here, the polymeric function $P$ is given by
\begin{equation} \label{polym_funct}
 P = \frac{\sqrt{1+\epsilon^{2}} - 1}{\sqrt{1+\epsilon^{2}} +1} ,
\end{equation}
where $\epsilon=\gamma \delta_b$, $\gamma$ is the Barbero-Immirzi parameter, and $\delta_b$ is polymeric parameter used for the quantization in LQG. Also, in the Eq. \eqref{lqgbh_metric} appears the parameter $a_0$ defined by
\begin{equation}
a_{0} = \frac{A_{\text{min}}}{8\pi},
\end{equation}
with $A_{\text{min}}$ being the minimal area in the context of LQG.

It is important to notice, that the Eq. \eqref{lqgbh_metric} is written in terms of the SDBH mass $m$ that is associated with the ADM mass as follows
\begin{equation}
M = m(1 + P)^{2} .
\end{equation}
Also, using the multiplicative factor of the angular part of Eq. \eqref{lqgbh_metric}, we can define a new radial coordinate
\begin{equation}
\label{rad_coord}
	R = \sqrt{r^2 + \frac{a_0^2}{r^2}}.
\end{equation}
Here, $R$ measures the circumference distance and is equal to the $r$ coordinate only in the asymptotic limit. Furthermore, from the Eq. \eqref{rad_coord}, we can see an important characteristic of the internal structure of the SDBH. When $r$ decreases from infinity to zero, the $R$ coordinate decreases from infinity to $R=\sqrt{2a_0}$ in $r=\sqrt{a_0}$, and then increases again to infinity. Considering the Eq. \eqref{rad_coord} in the external event horizon, i.e., in $r=r_{+}$, we get
\begin{equation} \label{r-horizon}
 R_{+} =  \sqrt{(2m)^{2} + \Big(\frac{a_{0}}{2m}\Big)^{2}} .
\end{equation}

\begin{figure}[h!]
\centering
\includegraphics[scale=0.7]{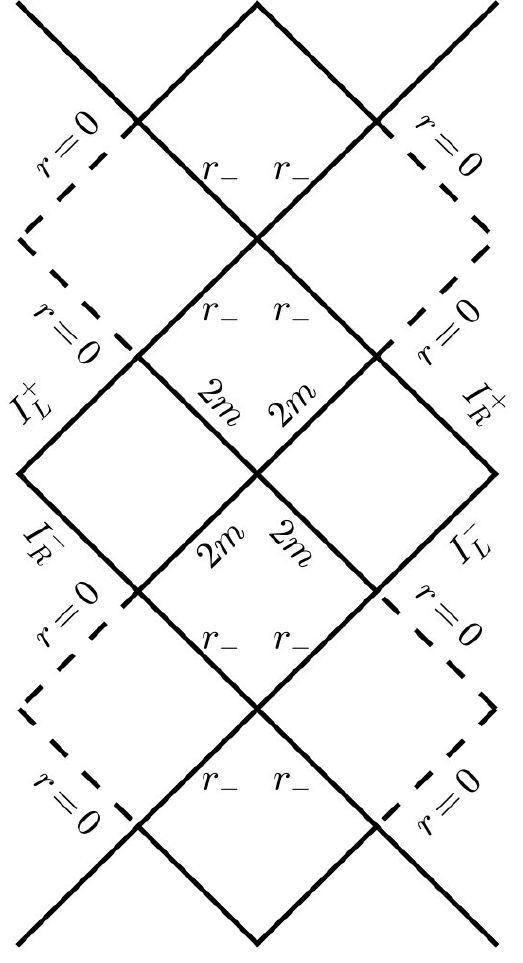}
\caption{Carter-Penrose diagram of the SDBH metric. The diagram has two asymptotic regions, being one at infinity and the other near the origin, where none observer can reach considering a finite time.}
\label{diag_bngql}
\end{figure}

A very interesting characteristic of this scenario is the self-duality of the metric in Eq. \eqref{lqgbh_metric}. The self-duality means that, if we introduce new coordinates, $\tilde{r}=a_{0}/r$ and $ \tilde{t}=tr_{*}^{2}/a_{0}$, the form of metric is preserved. The dual radial coordinate is given by $\tilde{r}=\sqrt{a_0}$ and corresponds to a minimal element of surface. Furthermore, the Eq. \eqref{rad_coord} can be written in the form $R=\sqrt{r^2+\tilde{r}^2}$  that clearly shows an asymptotically flat space, that is, a Schwarzschild region in the place of singularity in the limit as $r$ tends to zero. This region corresponds to a wormhole with the size of the order of the Planck length. The Carter-Penrose diagram for the SDBH is shown in Fig. \ref{diag_bngql}.

\section{Scalar perturbations and effective potential}
\label{effect_potent}

In this section, we will study a simple way to introduce the black holes perturbations, called scalar perturbations. Here, we shall consider a massive scalar field coupled to the Ricci scalar that is associated with background geometry.

The scalar field dynamics is described by the modified Klein-Gordon equation, given by
\begin{equation} \label{kg_eq}
	\begin{aligned}
		\left[ \frac{1}{\sqrt{-g}} \partial_{\mu} \left(g^{\mu \nu} \sqrt{-g} \partial_{\nu} \right) - \mu^2 - \xi \mathcal{R} \right] \Phi(t, r, \theta, \phi) = 0 .
	\end{aligned}
\end{equation}
Here, $\mu$ is the mass of the scalar field $\Phi$ and $\mathcal{R}$ is the Ricci scalar. The nonminimal coupling, $\xi$, have a similar form to the models used in cosmological context \cite{Abreu:1994fd, Mannheim:2005bfa, Mannheim:2009qi, Artymowski:2012is}. A conformal or Weyl transformation is the rescaling of the metric and the scalar field given by
\begin{equation} \label{conf_trans}
	\begin{aligned}
		\tilde{g}_{\mu \nu}(x) = \Omega(x)^2 g_{\mu \nu}(x) \ \ \ \ \ \text{and} \ \ \ \ \ \tilde{\Phi}(x) = \Omega(x)^{-1} \Phi(x) ,
	\end{aligned}
\end{equation}
where $\Omega(x)$ is a real, continuous, finite and non-vanishing function. In special, two values of $\xi$ are interesting: the $\xi=0$ called the minimally coupled case, and the conformally coupled case, $\xi=1/6$. It is possible to show that, if $\mu=0$ and $\xi=1/6$ the field equation given by Eq. \eqref{kg_eq} is invariant under conformal transformations (conformal symmetry) of Eq. \eqref{conf_trans} \cite{Mannheim:2011ds}. Also, in the Schwarzschild limit ($P=0$ and $a_0=0$) the Ricci scalar vanishes, and, consequently, we have the minimally coupled scalar field propagating in the Schwarzschild background.

So, we will start by developing the Eq. \eqref{kg_eq} considering the background metric given by Eq. \eqref{lqgbh_metric}. After performing some calculations, we get
\begin{equation} \label{eq1}
	\begin{aligned}
		& - \frac{a_{0}^2+r^4}{(r-r_{-}) (r-r_{+})(r+r_{*})^2} \frac{\partial^2 \Phi}{\partial t^2} + \frac{r^4}{\left(a_{0}^2+r^4\right) (r+r_{*})^2} \frac{\partial}{\partial r} \left[(r-r_{-}) (r-r_{+}) \frac{\partial \Phi}{\partial r}\right] \\ & + \frac{r^2}{a_{0}^2+r^4} \frac{1}{\text{sin}\theta} \frac{\partial}{\partial \theta} \left(\text{sin}\theta \frac{\partial \Phi}{\partial \theta}\right) + \frac{r^2}{a_{0}^2+r^4} \frac{1}{\text{sin}^2 \theta} \frac{\partial^2 \Phi}{\partial \phi^2} = \left(\mu^2 + \xi \mathcal{R} \right) \Phi ,
	\end{aligned}
\end{equation}
where Ricci scalar, is given by
\begin{equation} \label{ricci_scalar}
	\begin{aligned}
		\mathcal{R} &= g^{\mu \nu} \mathcal{R}_{\mu \nu} \\
		&= \frac{2 r^2}{\left(a_{0}^2+r^4 \right)^3 (r+r_{*})^4} \Bigg[- r_{*}^2 \left(2 a_{0}^2 r^4 \left(3 r^2+3 r (r_{-}+r_{+})-4 r_{-} r_{+}\right) + a_{0}^4 \left(6 r^2-r_{-} r_{+}\right) \right. \\ & \ \ \left. + r^8 \left(4 r^2+2 r (r_{-}+r_{+})-3
   		r_{-} r_{+}\right)\right)-r r_{*} \left(2 a_{0}^2 r^4 \left(-4 r^2+7 r (r_{-}+r_{+})-8 r_{-} r_{+}\right) \right. \\ & \left. \ \ + a_{0}^4 (3 r (r_{-}+r_{+})-4 r_{-} r_{+})+r^8 (3 r (r_{-}+r_{+})-4 r_{-} r_{+}) \right)+2 a_{0}^2 r^2 \left(a_{0}^2+3 			r^4\right) (r-r_{-})  \\ & \ \ \times (r- r_{+}) + r_{-}^4 \left(-\left(a_{0}^2+r^4\right)^2\right)-4 r r_{*}^3 \left(a_{0}^2+r^4\right)^2 \Bigg] .
	\end{aligned}
\end{equation}

\begin{figure}[h!]
\centering
\includegraphics[scale=0.5]{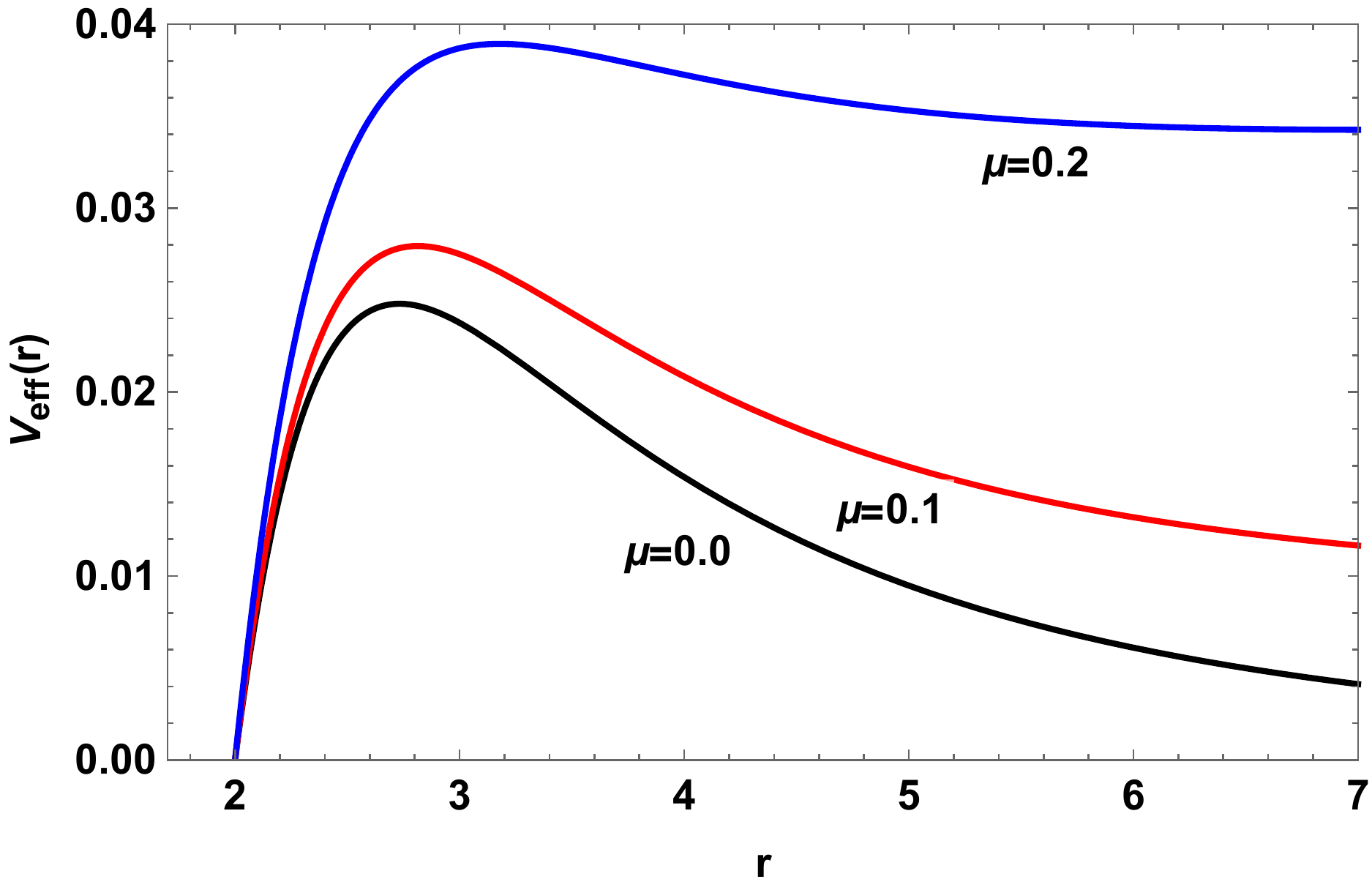}
\caption{Graph of the effective potential. This plot show the behavior of $V_{\text{eff}}(r)$ for different values of mass of scalar field: $\mu=0.0, 0.1$ and $0.2$. Here, we assume the constant values $m=1$, $\xi=0.1$, $a_0=\sqrt{3}/2$, $l=0$ and $P=0.1$.}
\label{fig_vr_mu}
\end{figure}

\begin{figure}[h!]
\centering
\includegraphics[scale=0.5]{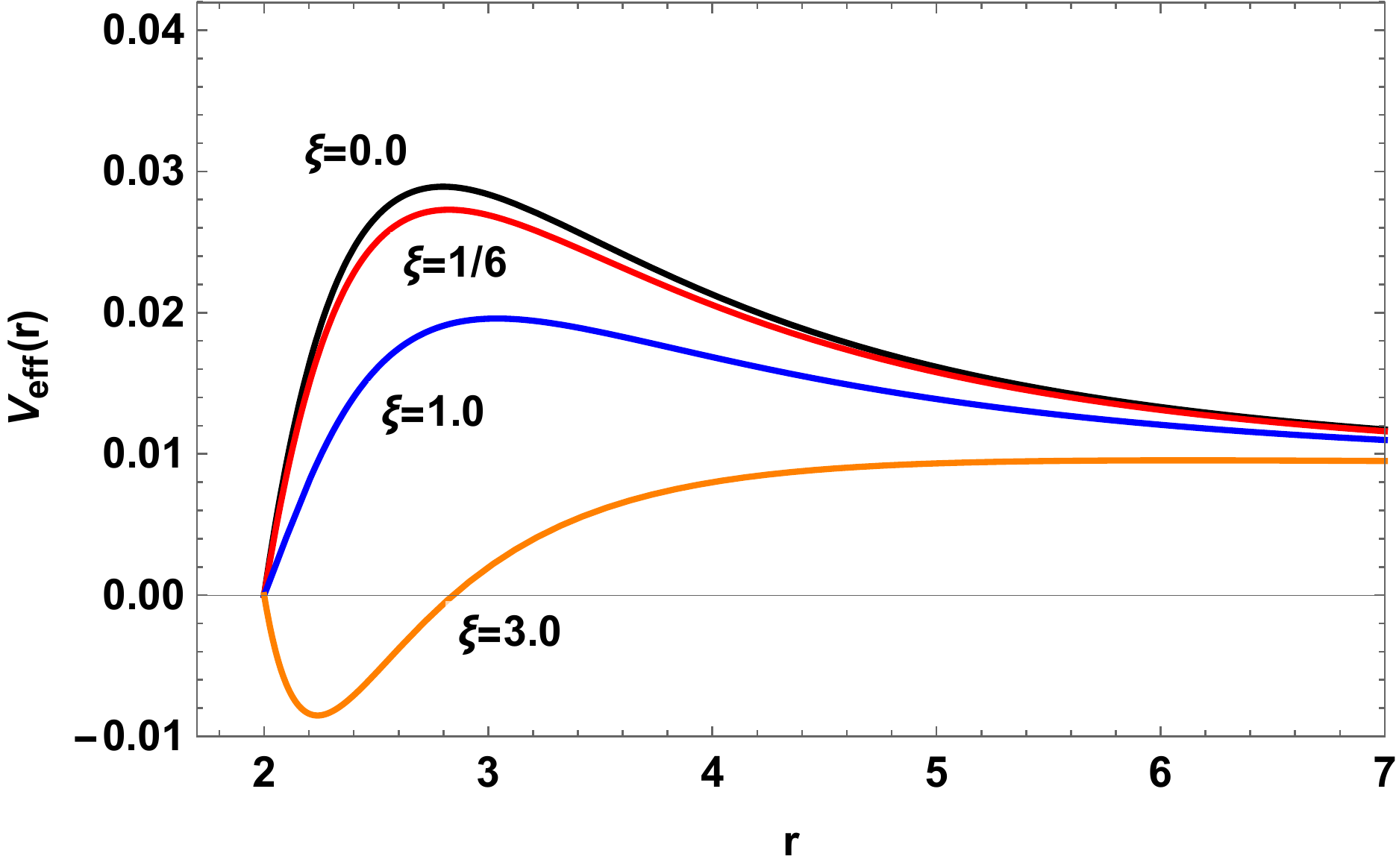}
\caption{Graph of the effective potential. This plot shows the behavior of $V_{\text{eff}}(r)$ for different values of the nonminimal coupling: $\xi=0.0, 1/6, 1.0$ and $3.0$. Here, we assume the constant values $m=1$, $\mu=0.1$, $a_0=\sqrt{3}/2$, $l=0$ and $P=0.1$.}
\label{fig_vr_xi}
\end{figure}

Now, we should introduce the scalar field through standard Ansatz in the following form:
\begin{equation} \label{ans_eq}
	\begin{aligned}
		\Phi(t, r, \theta, \varphi) = \psi(t, r) Y_{l}^{m}(\theta, \varphi) ,
	\end{aligned}
\end{equation}
such that, substituting in Eq. \eqref{eq1}, we find
\begin{equation} \label{eq2}
	\begin{aligned}
		& - \frac{\partial^2 \psi}{\partial t^2} + \frac{r^4 (r-r_{-}) (r-r_{+})}{\left(a_{0}^2+r^4\right)^2} \frac{\partial}{\partial r} \left[ (r-r_{-}) (r-r_{+}) \frac{\partial \psi}{\partial r} \right] \\ & - \frac{r^2 (r-r_{-}) (r-r_{+}) (r+r_{*})^2 l(l+1)}{\left(a_{0}		^2+r^4\right)^2} \psi = \frac{(r-r_{-}) (r-r_{+}) (r+r_{*})^2}{a_{0}^2+r^4} (\mu^2 + \xi \mathcal{R}) \psi .
	\end{aligned}
\end{equation}
Here, the $Y_{l}^{m}(\theta, \phi)$ denotes the spherical harmonics functions \cite{abramowitz1988handbook} that satisfy the relation:
\begin{equation}
	\begin{aligned}
		\frac{1}{\text{sin} \theta} \frac{\partial}{\partial \theta} \left( \text{sin} \theta \frac{\partial Y_{l}^{m}}{\partial \theta} \right) + \frac{1}{\text{sin}^{2} \theta} \frac{\partial^{2} Y_{l}^{m}}{\partial \phi^{2}} = - l(l+1) Y_{l}^{m} .
	\end{aligned}
\end{equation}
In this point, we can redefine the function $\psi$ as:
\begin{equation} \label{eq3}
	\begin{aligned}
		\psi = \frac{r}{\sqrt{a_{0}^2+r^4}} \Psi(r) e^{- i \omega t} ,
	\end{aligned}
\end{equation}
where the $\omega$ parameter denotes the quasinormal modes. Also, we should introduce the tortoise coordinate, $x$, through the relation:
\begin{equation} \label{eq_int_x}
	\begin{aligned}
		\frac{d}{d x} = \frac{r^2 (r-r_{-}) (r-r_{+})}{a_{0}^2+r^4} \frac{d}{d r} ,
	\end{aligned}
\end{equation}
that performing the integration of Eq. \eqref{eq_int_x}, we get
\begin{equation} \label{tort_coord}
	\begin{aligned}
		x = & r - \frac{a_{0}^2}{r r_{-} r_{+}} + \frac{ a_{0}^2 \left(r_{-} + r_{+}\right) \log (r)}{r_{-}^2 r_{+}^2}+\frac{\left(a_{0}^2+r_{-}^4\right) \log (r-r_{-})}{r_{-}^2 (r_{-}-r_{+})} \\ & - \frac{\left(a_{0}^2+r_{+}^4\right) \log (r-r_{+})}{r_{+}^2 				(r_{-}-r_{+})} .
	\end{aligned}
\end{equation}
Finally, substituting the Eqs. \eqref{ricci_scalar}, \eqref{eq3} and \eqref{eq_int_x} into Eq. \eqref{eq2}, we obtain a Schrödinger-like wave equation for the scalar perturbation of SDBH spacetime in the following form:
\begin{equation}
	\begin{aligned} \label{schrod_like_eq}
		\frac{d^2 \Psi}{d x^2} + \left[ \omega^2 - V_{\text{eff}}(r) \right] \Psi = 0 ,
	\end{aligned}
\end{equation}
where the effective potential is given by:
\begin{equation} \label{effec_pot}
	\begin{aligned}
		V_{\text{eff}}(r) & = \frac{(r-r_{-})(r-r_{+})}{\left(a_{0}^2+r^4\right)^4 (r+r_{*})^2} \Bigg[l (l+1) r^2 \left(a_{0}^2+r^4\right)^2 (r+r_{*})^4 + a_{0}^2 r^6 \left(4 \xi  \left(-r_{*}^2 \left(3 r^2+3 r \right. \right. \right. \\ & \ \ \left. \left. 		\left. \times (r_{-}+r_{+})-4 r_{-} r_{+}\right)+r r_{*} \left(4 r^2-7 r (r_{-}+r_{+})+8 r_{-} r_{+}\right)+3 r^2 (r-r_{-}) (r-r_{+})-4 r r_{*}^3 \right. \right. \\ & \ \ \left. \left. -r_{*}^4\right)+3 \mu ^2 r^2 (r+		r_{*})^4\right)+a_{0}^4 r^2 \left(2 \xi \left(r_{*}^2 \left(r_{-} r_{+}-6 r^2\right)+2 r^2 (r-r_{-}) (r-r_{+}) \right. \right. \\ & \ \ \left. \left. +r r_{*} (4 r_{-} r_{+}-3 r (r_{-}+r_{+}))-4 r r_{*}^3-r_{*}^4\right)+3 \mu^2 r^2 (r+r_{*})^4\right)+r^3 (r+r_{*})^2 \left(10 a_{0}^2 r^3 \right. \\ & \ \ \left. \times (r-r_{-}) (r-r_{+})+a_{0}^4 (-2 r+r_{-}+r_{+})+r^7 (r (r_{-}+r_{+})-2 r_{-} r_{+})\right)+a_{0}^6 \mu^2 (r+r_{*})^4 \\ & \ \ -2 \xi  r^{10} r_{*} \left(r^2 (3 (r_{-}+r_{+})+4 r_{*})+2 r (r_{-} (r_{*}-2 r_{+})+r_{*} (r_{+}+2 r_{*}))-3 r_{-} r_{+} r_{*}+r_{*}^3\right) \\ & \ \ +\mu^2 r^{12} (r+r_{*})^4 \Bigg] .
	\end{aligned}
\end{equation}
For easy visualization of the behavior of effective potential of Eq. \eqref{effec_pot}, we have plotted the graphs shown in Figs. \ref{fig_vr_mu} and  \ref{fig_vr_xi}. So, we can see from Fig. \ref{fig_vr_mu} that in the asymptotic limit, i.e. $r \longrightarrow \infty$, the effective potential exhibits the following behavior $V_{\text{eff}}(r) \sim \mu^{2}$. Also, it is observed that for small masses of the scalar field, the effective potential still has the form of barrier potential. However, by increasing values of $\mu$, the form of the effective potential is turned a potential barrier into a step potential.
In Fig. \ref{fig_vr_xi}, we can observe that for large couplings values (orange curve) we have an instability phase, where the effective potential is negative. Also, the limit case when $\xi \longrightarrow 0$  denoted by the black curve, is nearly coincident with the conformal gravity value ($\xi=1/6$). Due to these facts, we should consider parameters with small values so we can apply the WKB approximation correctly for calculation of QNMs.

\section{Quasinormal modes} 
\label{qnm_sdbh}

\begin{figure}[h!]
\centering
\includegraphics[scale=0.5]{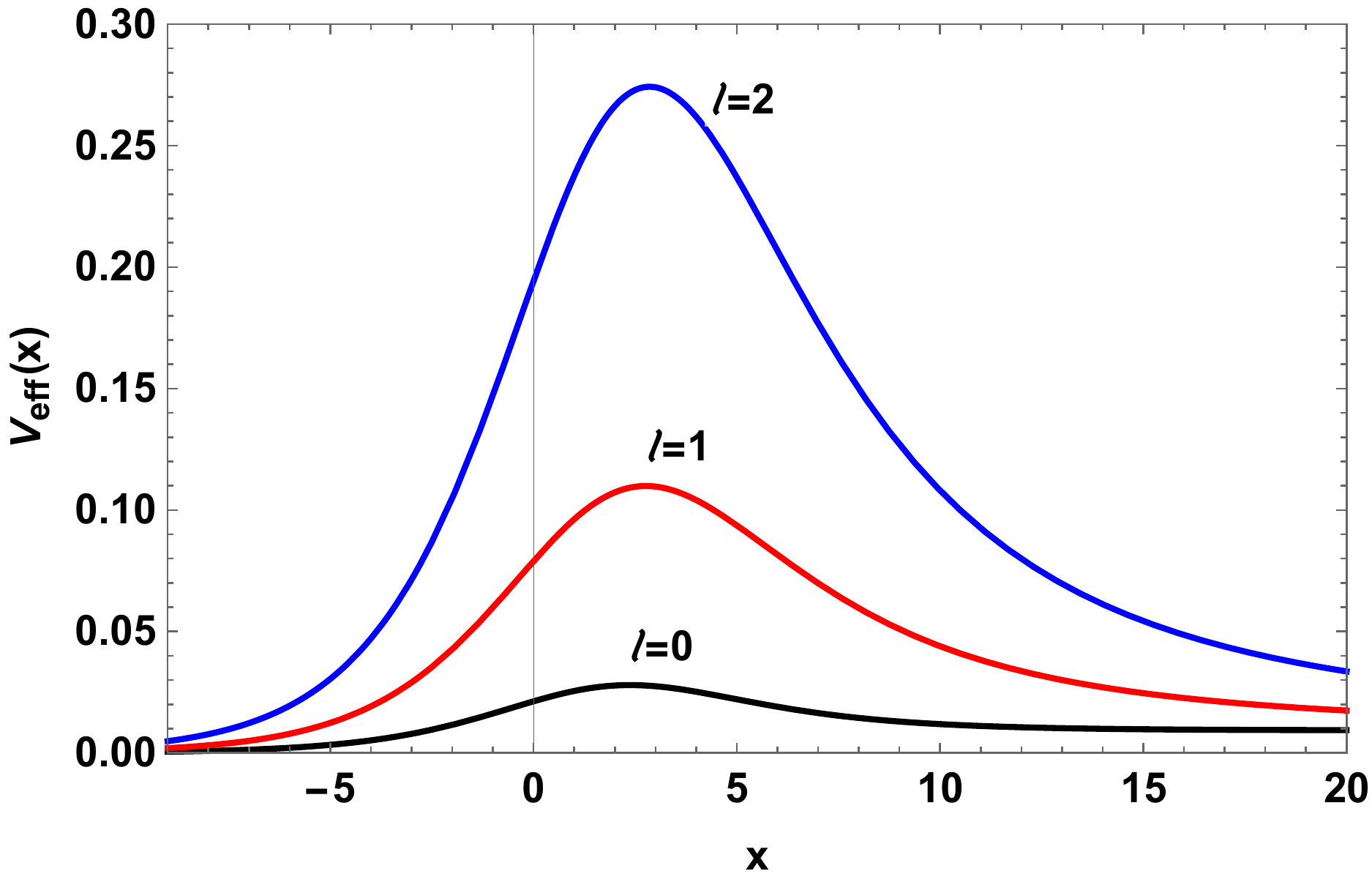}
\caption{Graph of the effective potential depending on tortoise coordinate. This plot shows the behavior of $V_{\text{eff}}(x)$ for different values of the multipole quantum number: $l=0, 1$ and $2$. Here, we assume the constant values $m=1$, $\mu=0.1$, $\xi=0.1$, $a_0=\sqrt{3}/2$ and $P=0.1$.}
\label{fig_vx_l}
\end{figure}

\begin{figure}[h!]
\centering
\includegraphics[scale=0.5]{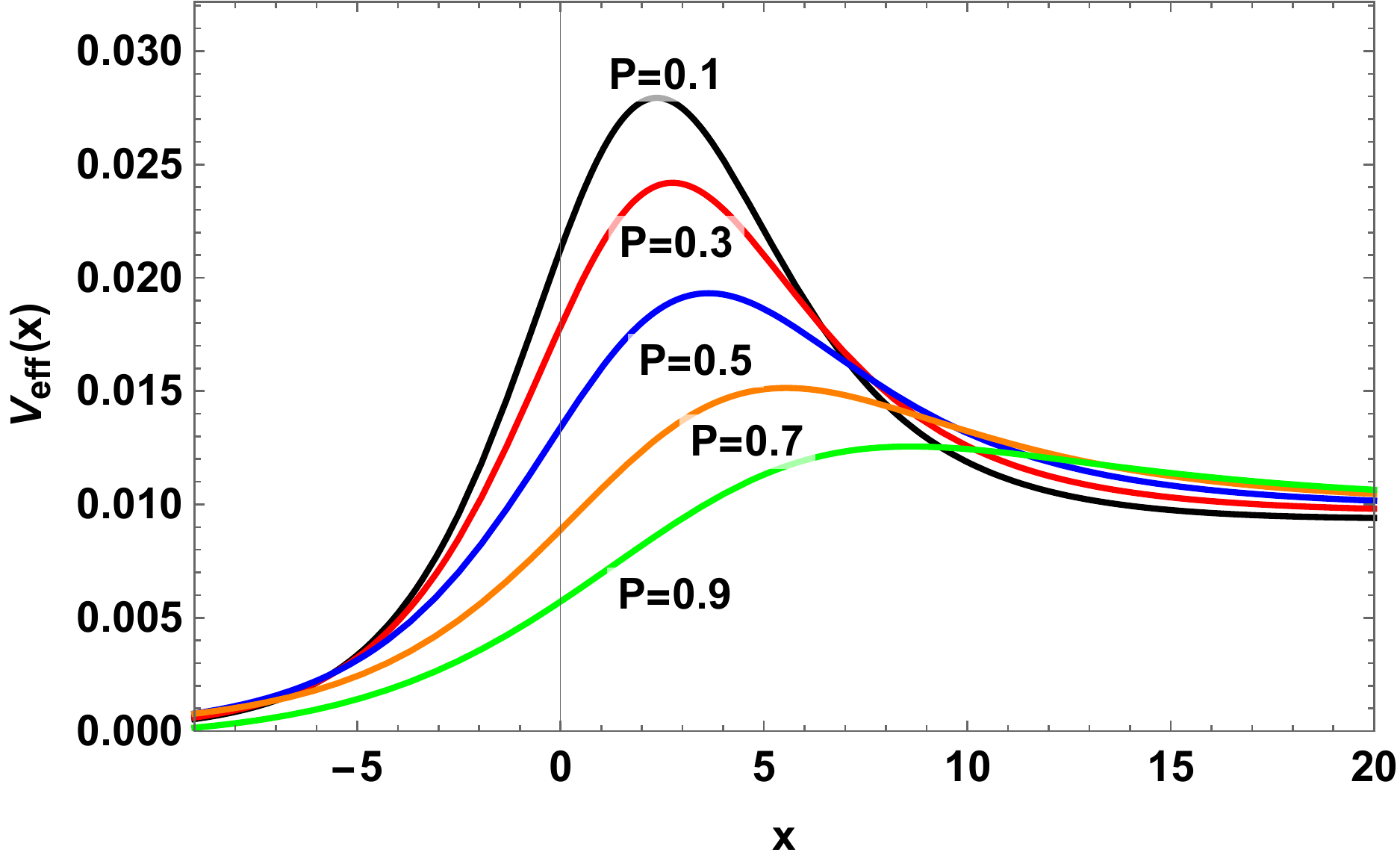}
\caption{Graph of the effective potential depending on tortoise coordinate. This plot shows the behavior of $V_{\text{eff}}(x)$ for different values of the polymeric parameter: $P=0.1, 0.3, 0.5, 0.7$ and $0.9$. Here, we assume the constant values $m=1$, $\mu=0.1$, $\xi=0.1$, $a_0=\sqrt{3}/2$ and $l=0$.}
\label{fig_vx_p}
\end{figure}

In this section, we will focus on the computation of the QNMs for the quantum black hole described through metric of the Eq. \eqref{lqgbh_metric}. As we saw in the last section, after considering the scalar perturbations to SDBH spacetime we can find a Schrödinger-like equation given by Eq. \eqref{schrod_like_eq} with an effective potential given by Eq. \eqref{effec_pot}. Also, considering the effective potential depending on tortoise coordinate of Eq. \eqref{tort_coord}, the $V_{\text{eff}}(x)$ assumes constant values near event horizon ($x=-\infty$) and at infinity ($x=\infty$) and has a maximum value at some intermediate point ($x=x_0$). This behavior is shown in Figs. \ref{fig_vx_l} and \ref{fig_vx_p}, where we can see an increase of peak of the effective potential with the $l$ values, and a decrease with $P$ values.
\

As earlier mentioned, the QNMs are complex frequencies and can be expressed in the following form:
\begin{equation}
\begin{aligned}
\omega=\omega_{\text{R}}+i\omega_{I} ,
\end{aligned}
\end{equation}
where the real part ($\omega_{\text{R}}$) determines the normal frequency of the oscillations, while the imaginary part ($\omega_{I}$) represents the damping time of vibration modes. Also, we can get information about the stability of BHs from the analysis of QNMs. The BHs are unstable when $\omega_{\text{I}} > 0$ and stable when $\omega_{\text{I}} < 0$.

Due to behavior of $V_{\text{eff}}(x)$ showed in Fig. \ref{fig_vx_l}, we can make a direct analogy with the problem of scattering near the peak of the barrier potential of quantum mechanics, where $\omega^2$ in Eq. \eqref{schrod_like_eq} plays the role of the energy. Several methods to compute the QNMs have been developed \cite{Berti:2009kk, Kokkotas:1999bd, Nollert:1999ji, Konoplya:2004ip}. However, we chose to apply an approximated method, which is the well-known WKB approach introduced by Schutz and Will \cite{Schutz:1985km}. This treatment was later improved to the 3rd order by Iyer and Will \cite{Iyer:1986np}, and is applied in our calculus. Thus, the QNMs ($\omega=\omega_{n}$) that appear in Eq. \eqref{schrod_like_eq} are determined by the following equation:
\begin{equation} 
 \label{qnm_eq}
 \omega_{n}=\sqrt{\left(V_{0} + \Delta \right) - i \left(n + \frac{1}{2}\right) \sqrt{-2 V^{\prime \prime}_{0}} \left(1 + \Omega \right)} ,
\end{equation}
where
\begin{equation}
 \Delta = \frac{1}{8}\left(\frac{V_{0}^{(4)}}{V^{\prime \prime}_{0}}\right)
 \left(\frac{1}{4}+\alpha^{2}\right)-\frac{1}{288}\left(\frac{V^{\prime \prime \prime}_{0}}{V^{\prime \prime}_{0}}\right)^{2}\left(7+60\alpha^{2}
 \right) \ ,
\end{equation}
\begin{equation}
 \begin{aligned}
 \Omega &= -\frac{1}{2V^{\prime \prime}_{0}} \left \{ \frac{5}{6912}\left(\frac{V^{\prime \prime \prime}_{0}}{V^{\prime \prime}_{0}}\right)^{4}\left(77+188\alpha^{2}
 \right) - \frac{1}{384}\left[\frac{\left(V^{\prime \prime \prime}_{0}\right)^{2}\left(V^{(4)}_{0}\right)}{\left(V^{\prime \prime}_{0}\right)^{3}}\right]\left(51+100\alpha^{2}
 \right) \right. \\ & \ \ \ \left. \right. + \frac{1}{2304}\left(\frac{V^{(4)}_{0}}{V^{\prime \prime}_{0}}\right)^{2}\left(65+68\alpha^{2}\right) + 
 \frac{1}{288}\left(\frac{V^{\prime \prime \prime}_{0}V^{(5)}_{0}}{\left(V^{\prime \prime}_{0}\right)^{2}}\right)\left(19+28\alpha^{2}\right) \\ & \ \ \ \left. - \frac{1}{288}\left(\frac{V^{(6)}_{0}}{V^{\prime \prime}_{0}}\right)\left(5+4\alpha^{2}\right) \right \} \ .
 \end{aligned}
\end{equation}
Here, we have $\alpha=n+1/2$ and $V_0^{(n)}$ denotes the $n$-order derivative of the effective potential on the maximum point $x_0$. Other developments recent have extended the WKB method of 3th to 6th order \cite{Konoplya:2003ii}, well as to higher-orders \cite{Matyjasek:2017psv, Konoplya:2019hlu}. However, these approaches of higher orders should reproduce results very nearly of results reproduced in the third order, in cases that $n \leqslant l$. So, as we have interesting in small overtones number, $n$, the 3rd order WKB method has good accuracy to QNMs.

\begin{figure}
    \centering
    \subfloat[\centering Real part]{{\includegraphics[width=7cm]{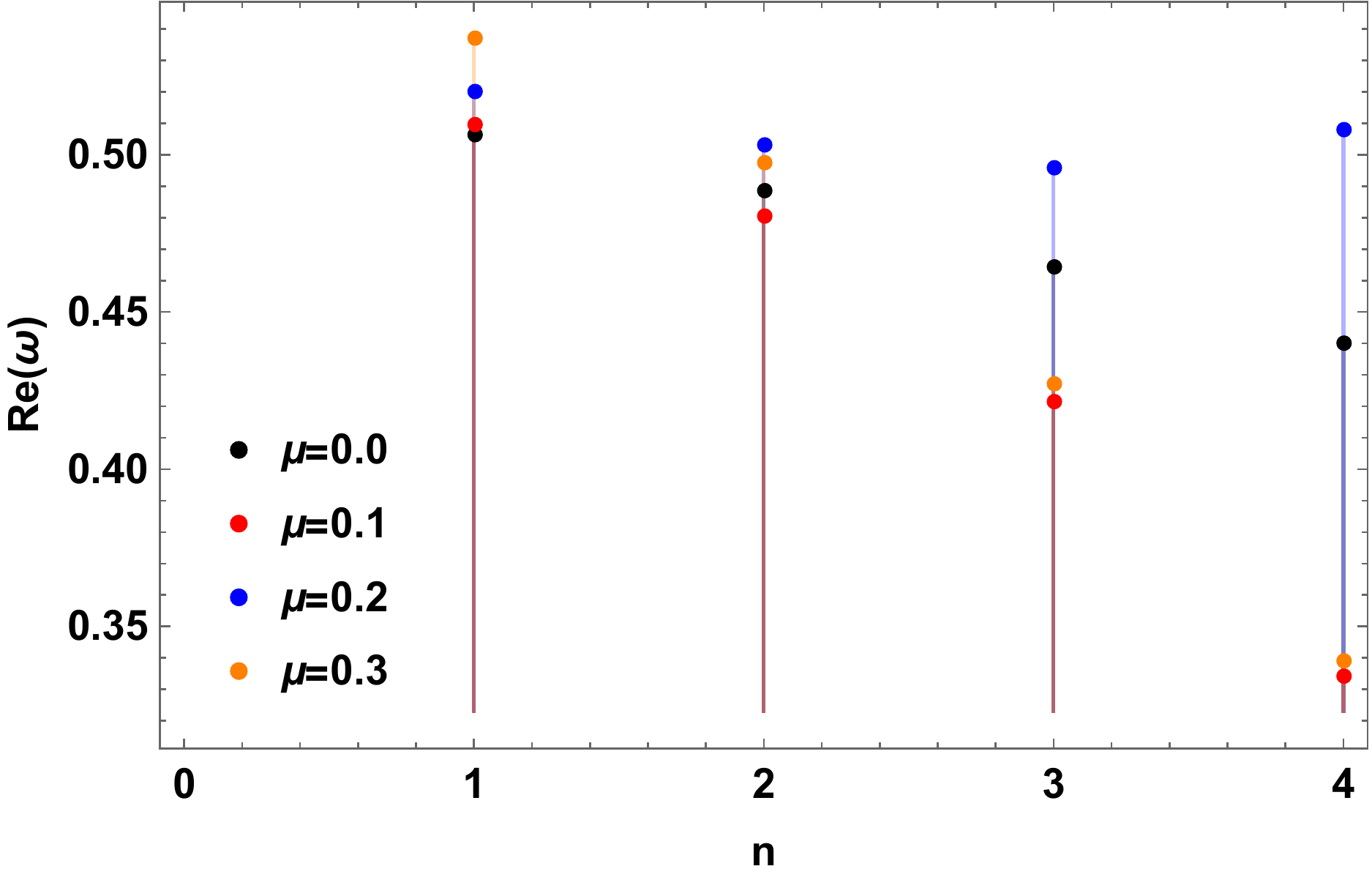} }}
    \qquad
    \subfloat[\centering Imaginary part]{{\includegraphics[width=7cm]{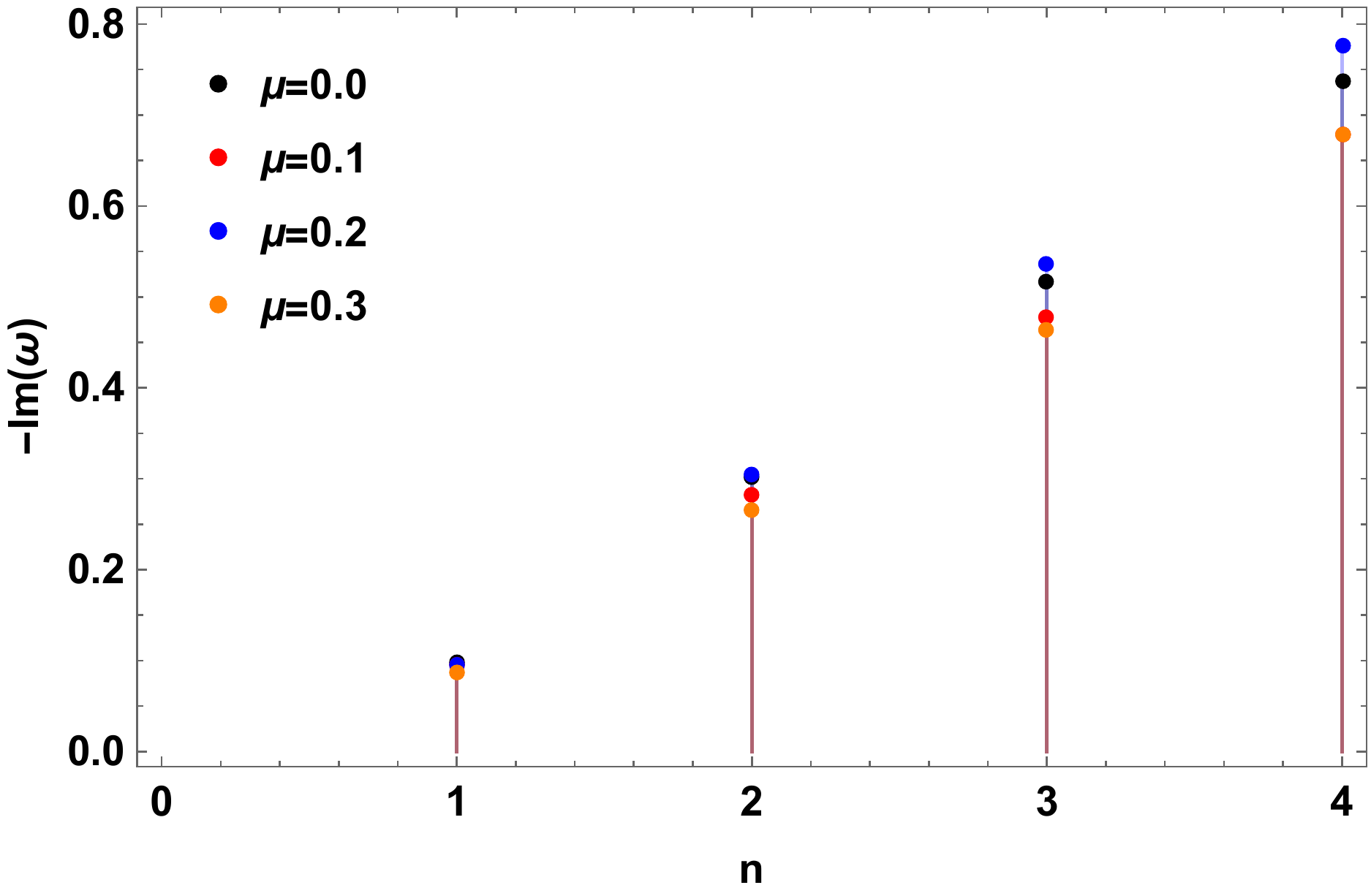} }}
    \caption{Graphs for the behavior of the QNMs considering $l=2$ and $\mu=0.0, 0.1, 0.2 \ \text{and} \ 0.3$. In plot (a) is shown the real part, while the imaginary part is shown in (b).}
    \label{fig_qnm_l2_mu}
\end{figure}

\begin{figure}
    \centering
    \subfloat[\centering Real part]{{\includegraphics[width=7cm]{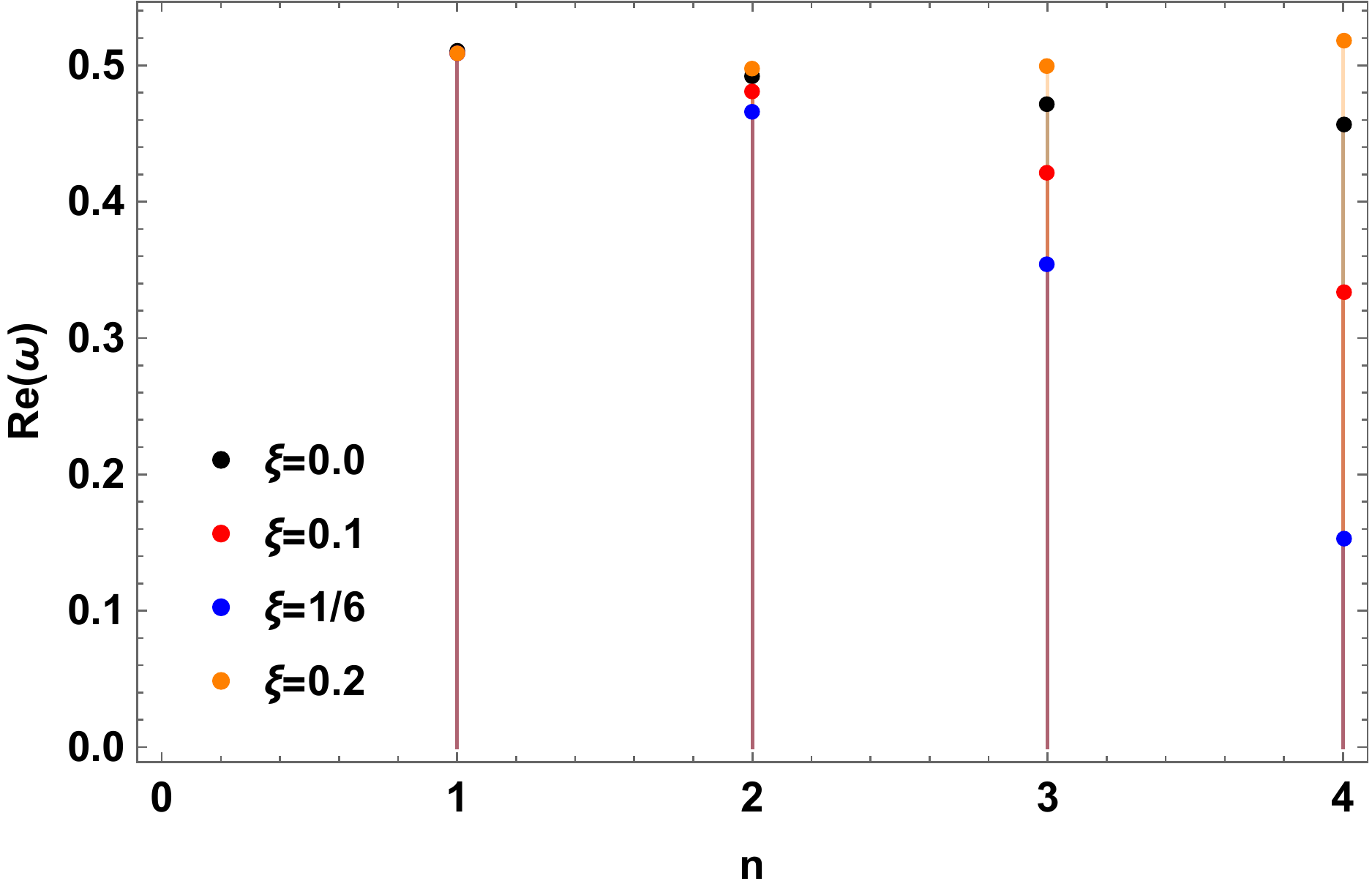} }}
    \qquad
    \subfloat[\centering Imaginary part]{{\includegraphics[width=7cm]{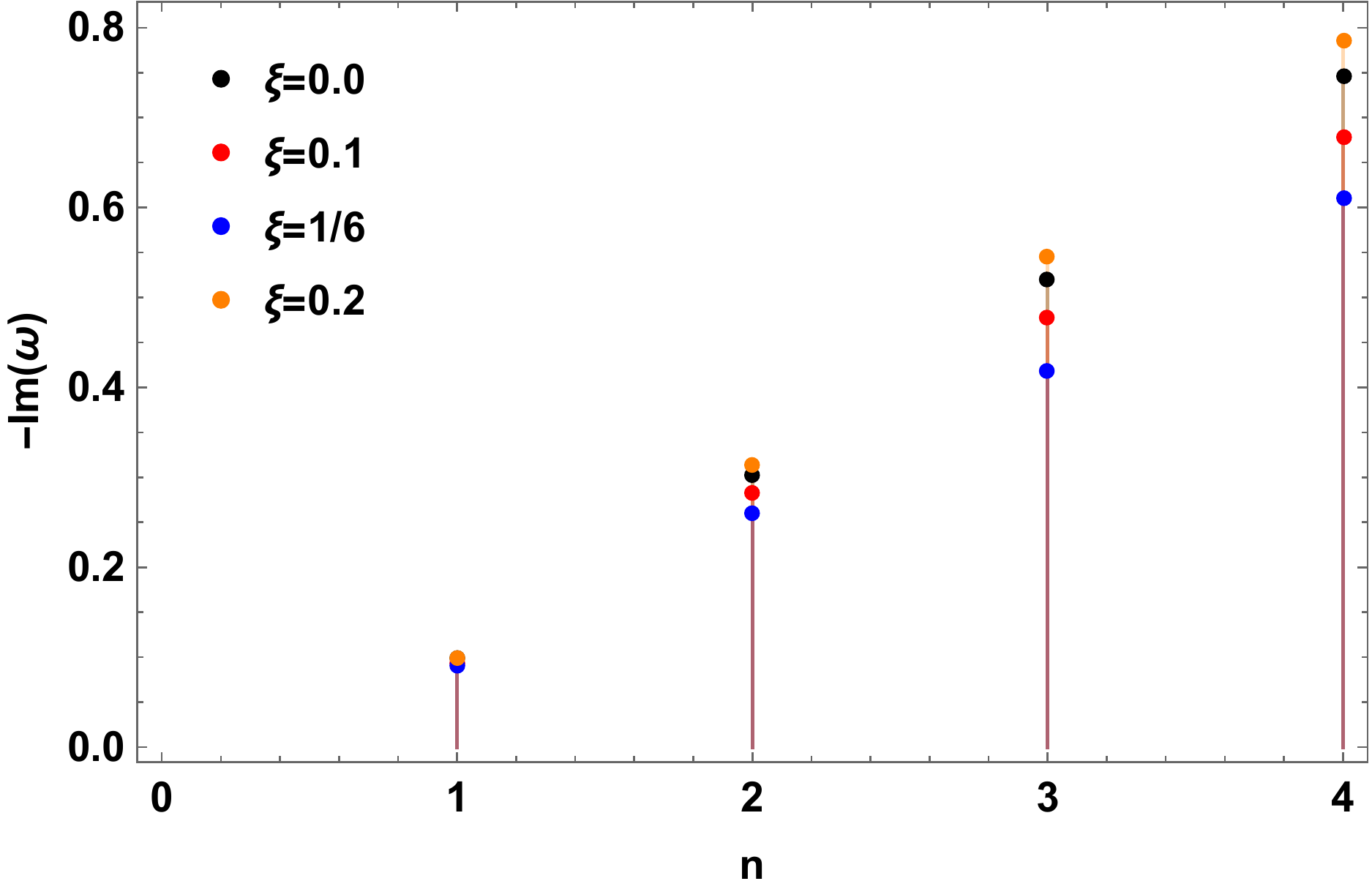} }}
    \caption{Graphs for the behavior of the QNMs considering $l=2$ and $\xi=0.0, 0.1, 1/6 \ \text{and} \ 0.2$. In plot (a) is shown the real part, while the imaginary part is shown in (b).}
    \label{fig_qnm_l2_xi}
\end{figure}

\begin{figure}
    \centering
    \subfloat[\centering Real part]{{\includegraphics[width=7cm]{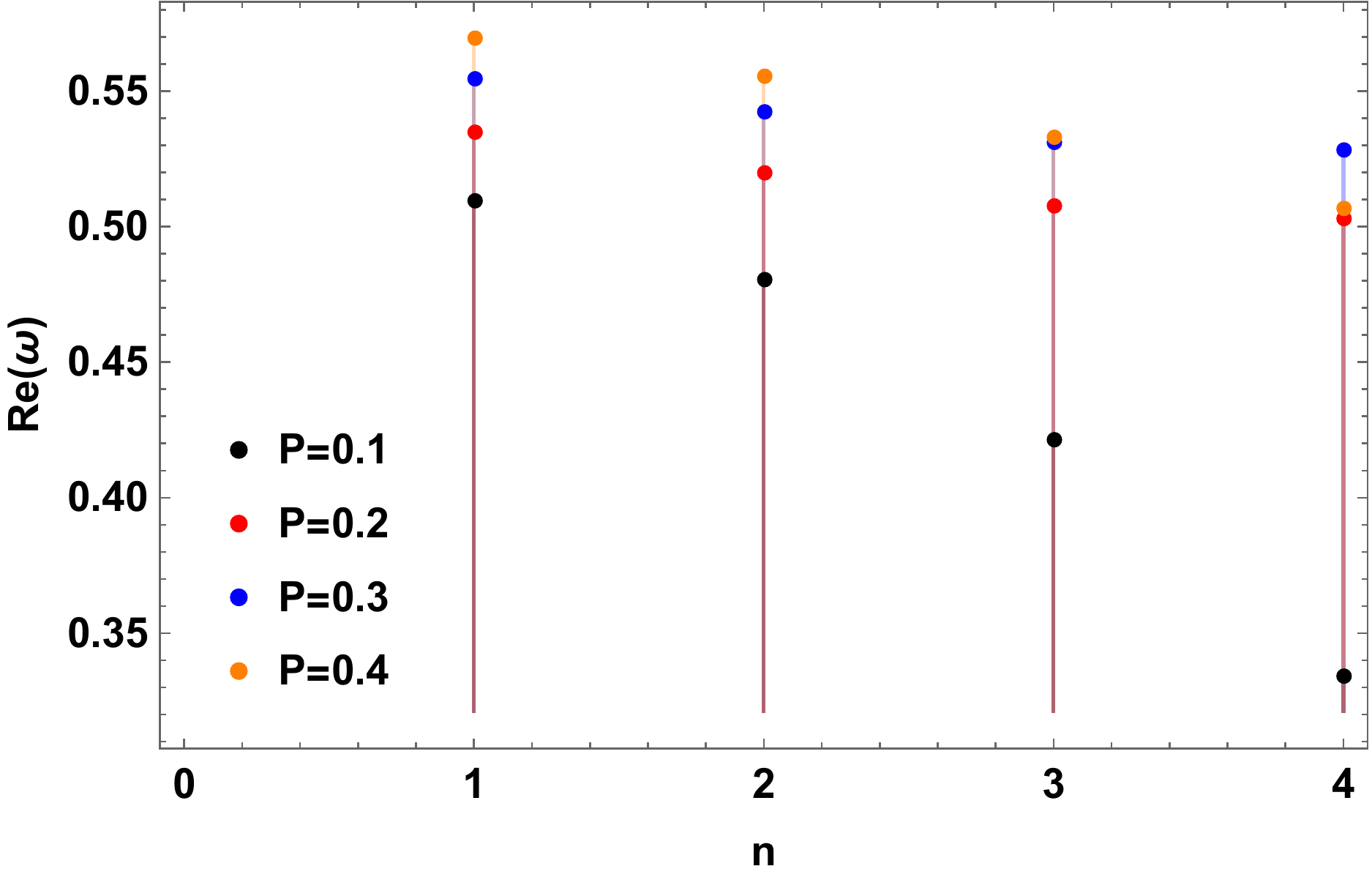} }}
    \qquad
    \subfloat[\centering Imaginary part]{{\includegraphics[width=7cm]{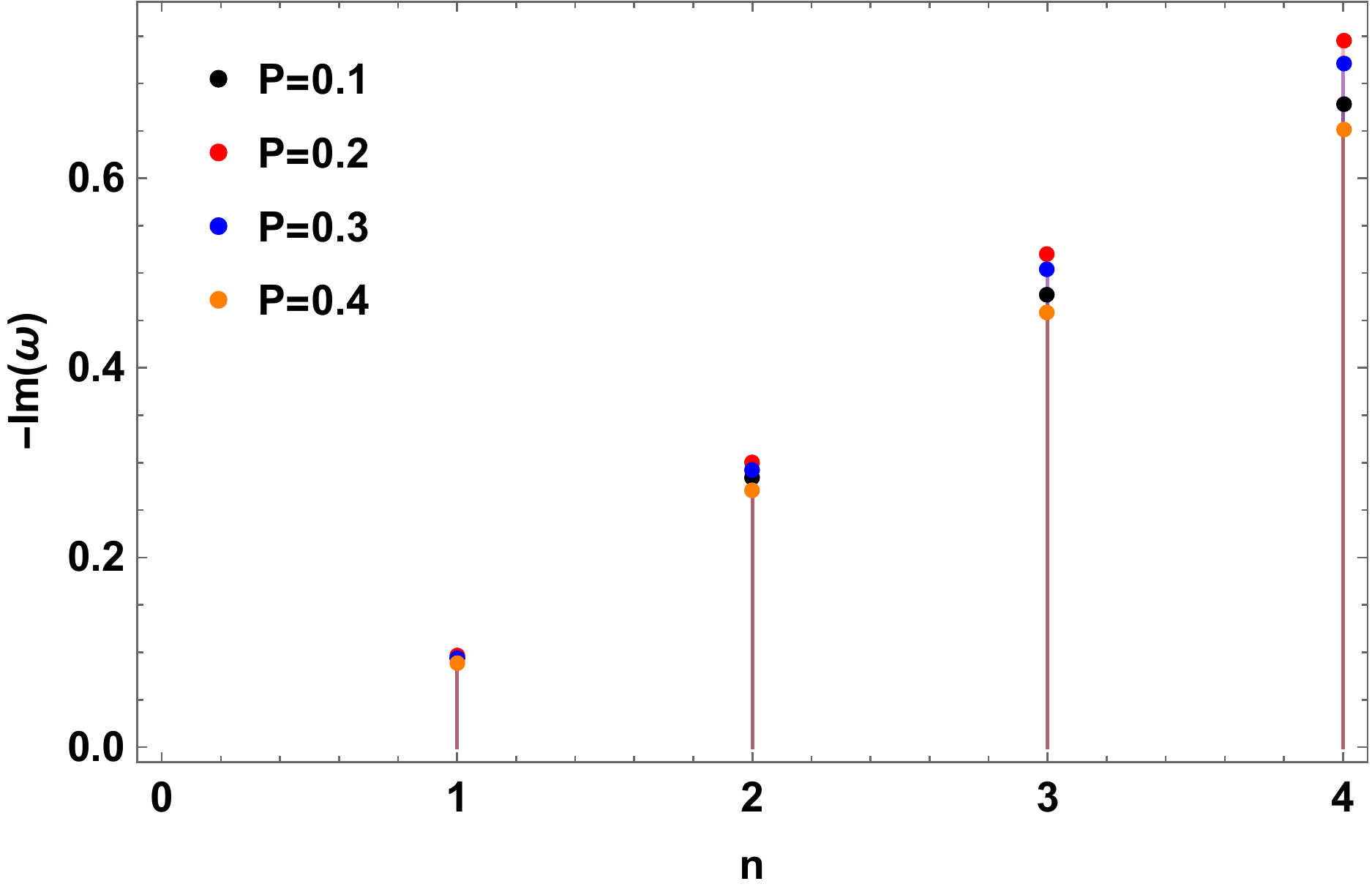} }}
    \caption{Graphs for the behavior of the QNMs considering $l=2$ and $P=0.1, 0.2, 0.3 \ \text{and} \ 0.4$. In plot (a) is shown the real part, while the imaginary part is shown in (b).}
    \label{fig_qnm_l2_p}
\end{figure}

Thus, using the Eq. \eqref{effec_pot} and Eq. \eqref{qnm_eq}, we can calculate the QNMs for the self-dual black hole considering different values of the parameters. In Tab. \ref{table1}, we show the fundamental QNMs ( i.e., $n=0$) for the case $l=0$ and considering different values of the other parameters. We can note that the QNMs values considering $\mu=0$, for the case $\xi=0$ are different from the conformal coupling case ($\xi=1/6$). Furthermore, for the case when polymeric $P$ and minimal area $a_0$ parameters tend to zero, the results converge to SchBH \cite{Simone:1991wn, Konoplya:2005hr}.
\begin{table}[!h]
\begin{center}
\scriptsize
\begin{tabular}{||c||c||c||c|c||c||c||c||c|}
\hline
{\bf $P$} & {\bf $\mu$} & {\bf $\xi$} & {\bf $\omega_{0}$} & {\ \ } & {\bf $P$} & {\bf $\mu$} & {\bf $\xi$} & {\bf $\omega_{0}$} \\
\hline
\hline
{\bf $0.1$} & {\bf $0.0$} & {\bf $0.0$} & {\bf $0.083846 - 0.106928 i$} & {} & {\bf $0.3$} & {\bf $0.0$} & {\bf $0.0$} & {\bf $0.072469 - 0.089438 i$} \\
\hline
{\bf $0.1$} & {\bf $0.0$} & {\bf $0.1$} & {\bf $0.116404 - 0.138260 i$} & {} & {\bf $0.3$} & {\bf $0.0$} & {\bf $0.1$} & {\bf $0.067228 - 0.103055 i$} \\
\hline
{\bf $0.1$} & {\bf $0.0$} & {\bf $1/6$} & {\bf $0.076978 - 0.110763 i$} & {} & {\bf $0.3$} & {\bf $0.0$} & {\bf $1/6$} & {\bf $0.024120 - 0.094200 i$} \\
\hline
{\bf $0.1$} & {\bf $0.0$} & {\bf $0.2$} & {\bf $0.098475 - 0.128178 i$} & {} & {\bf $0.3$} & {\bf $0.0$} & {\bf $0.2$} & {\bf $0.043868 - 0.106240 i$} \\
\hline
{\bf $0.1$} & {\bf $0.1$} & {\bf $0.0$} & {\bf $0.103053 - 0.111495 i$} & {} & {\bf $0.3$} & {\bf $0.1$} & {\bf $0.0$} & {\bf $0.090732 - 0.085852 i$} \\
\hline
{\bf $0.1$} & {\bf $0.1$} & {\bf $0.1$} & {\bf $0.034850 - 0.064386 i$} & {} & {\bf $0.3$} & {\bf $0.1$} & {\bf $0.1$} & {\bf $0.101386 - 0.111908 i$} \\
\hline
{\bf $0.1$} & {\bf $0.1$} & {\bf $0.2$} & {\bf $0.091391 - 0.111355 i$} & {} & {\bf $0.3$} & {\bf $0.1$} & {\bf $0.2$} & {\bf $0.038487 - 0.081088 i$} \\
\hline
\end{tabular}
\caption{Fundamental QNMs of SDBH considering the constant values: $m=1$, $l=0$ and $a_0=\sqrt{3}/2$.}
\label{table1}
\end{center}
\end{table}
Also, the QNMs for the SDBH considering the first overtone numbers ($n$) are shown in Tab. \ref{table2} for the case with $l=1$. We have adopted some typical values to the other parameters, for instance, $P=0.1, 0.3$ and $0.5$. However, for the case $\xi=0$ and $\mu=0$ the results agree with the values obtained in \cite{Santos:2015gja}. Finally, for better visualization of the effects due to the new parameters in the QNMs spectrum, we have shown the behavior by graphs in Figs. \ref{fig_qnm_l2_mu}, \ref{fig_qnm_l2_xi} and \ref{fig_qnm_l2_p}, where has been plotted the real and imaginary parts of $\omega$ for the case $l=2$.
\begin{table}[!h]
\begin{center}
\scriptsize
\begin{tabular}{||c||c||c||c|c|c|}
\hline
{\bf $P$} & {\bf $\mu$} & {\bf $\xi$} & {\bf $\omega_{0}$} & {\bf $\omega_{1}$} & {\bf $\omega_{2}$} \\
\hline
\hline
{\bf $0.1$} & {\bf $0.1$} & {\bf $0.0$} & {\bf $0.308677 - 0.097383 i$} & {\bf $0.280706 - 0.312954 i$} & {\bf $0.255548 - 0.542527 i$} \\
\hline
{\bf $0.1$} & {\bf $0.1$} & {\bf $0.1$} & {\bf $0.305982 - 0.093379 i$} & {\bf $0.262207 - 0.297952 i$} & {\bf $0.201681 - 0.520149 i$} \\
\hline
{\bf $0.1$} & {\bf $0.1$} & {\bf $0.2$} & {\bf $0.306023 - 0.098157 i$} & {\bf $0.281336 - 0.316308 i$} & {\bf $0.263790 - 0.548014 i$} \\
\hline
{\bf $0.1$} & {\bf $0.2$} & {\bf $0.0$} & {\bf $0.321783 - 0.082089 i$} & {\bf $0.249314 - 0.273727 i$} & {\bf $0.152563 - 0.505167 i$} \\
\hline
{\bf $0.1$} & {\bf $0.2$} & {\bf $0.1$} & {\bf $0.319751 - 0.079161 i$} & {\bf $0.234081 - 0.261550 i$} & {\bf $0.107235 - 0.494258 i$} \\
\hline
{\bf $0.1$} & {\bf $0.2$} & {\bf $0.2$} & {\bf $0.319462 - 0.083113 i$} & {\bf $0.252371 - 0.279548 i$} & {\bf $0.170198 - 0.512473 i$} \\
\hline
{\bf $0.3$} & {\bf $0.1$} & {\bf $0.0$} & {\bf $0.334209 - 0.090020 i$} & {\bf $0.299191 - 0.280891 i$} & {\bf $0.245795 - 0.485544 i$} \\
\hline
{\bf $0.3$} & {\bf $0.1$} & {\bf $0.1$} & {\bf $0.329608 - 0.089689 i$} & {\bf $0.294418 - 0.280507 i$} & {\bf $0.241676 - 0.485395 i$} \\
\hline
{\bf $0.3$} & {\bf $0.1$} & {\bf $0.2$} & {\bf $0.324837 - 0.088824 i$} & {\bf $0.287367 - 0.277780 i$} & {\bf $0.229716 - 0.481428 i$} \\
\hline
{\bf $0.3$} & {\bf $0.2$} & {\bf $0.0$} & {\bf $0.351419 - 0.083136 i$} & {\bf $0.304862 - 0.269468 i$} & {\bf $0.245882 - 0.482330 i$} \\
\hline
{\bf $0.3$} & {\bf $0.2$} & {\bf $0.1$} & {\bf $0.348755 - 0.088806 i$} & {\bf $0.325524 - 0.296933 i$} & {\bf $0.327314 - 0.530312 i$} \\
\hline
{\bf $0.3$} & {\bf $0.2$} & {\bf $0.2$} & {\bf $0.344568 - 0.088241 i$} & {\bf $0.320773 - 0.295941 i$} & {\bf $0.322452 - 0.529069 i$} \\
\hline
{\bf $0.5$} & {\bf $0.1$} & {\bf $0.0$} & {\bf $0.348520 - 0.082800 i$} & {\bf $0.333270 - 0.258464 i$} & {\bf $0.319151 - 0.445290 i$} \\
\hline
{\bf $0.5$} & {\bf $0.1$} & {\bf $0.1$} & {\bf $0.341353 - 0.082096 i$} & {\bf $0.325010 - 0.256477 i$} & {\bf $0.308893 - 0.442069 i$} \\
\hline
{\bf $0.5$} & {\bf $0.1$} & {\bf $0.2$} & {\bf $0.334788 - 0.084140 i$} & {\bf $0.327065 - 0.267313 i$} & {\bf $0.334821 - 0.464291 i$} \\
\hline
{\bf $0.5$} & {\bf $0.2$} & {\bf $0.0$} & {\bf $0.365596 - 0.068507 i$} & {\bf $0.310780 - 0.200578 i$} & {\bf $0.176663 - 0.346535 i$} \\
\hline
{\bf $0.5$} & {\bf $0.2$} & {\bf $0.1$} & {\bf $0.360323 - 0.074073 i$} & {\bf $0.327514 - 0.234962 i$} & {\bf $0.283478 - 0.415786 i$} \\
\hline
{\bf $0.5$} & {\bf $0.2$} & {\bf $0.2$} & {\bf $0.353556 - 0.072158 i$} & {\bf $0.315734 - 0.227651 i$} & {\bf $0.258431 - 0.403470 i$} \\
\hline
\end{tabular}
\caption{The First QNMs for the SDBH considering the constant parameters: $m=1$, $a_0=\sqrt{3}/2$ and $l = 1$.}
\label{table2}
\end{center}
\end{table}

\section{Concluding remarks}
\label{concluding}

The gravitational-wave observations have opened a new window to gravitational physics research. In this framework, the black holes offer a great scenario to test the predictions of candidates to quantum gravity theories. So, we have studied the black hole perturbations and the quasinormal modes spectrum to a quantized version of the Schwarzschild solution, which is known as a self-dual black hole \cite{Modesto:2009ve} and consists of a loop quantum gravity solution.

In present present work, we have considered massive scalar perturbations coupled with gravity through Ricci scalar. So, we can find a Schrödinger-like equation with an effective potential given by Eq. \eqref{effec_pot}. Then, we use the WKB approach to get the quasinormal modes showed in the tables \ref{table1}, \ref{table2}. We  consider different parameters values, so that we can get small corrections of classical solution. For instance, we consider $\mu=0.1, 0.2$, $\xi=0.1, 0.2$ and $P=0.1, 0.2, 0.5$. Also, we have considered the values to the angular number being $l=0,1$ and $2$. Due to limitations of the WKB approach, we have considered only small values to overtone number, $n$, \cite{Schutz:1985km, Iyer:1986np, Kokkotas:1999bd}. Therefore, for a better visualization of QNMs behavior, we plotted in Figs. \ref{fig_qnm_l2_mu}, \ref{fig_qnm_l2_xi} and \ref{fig_qnm_l2_p} the behavior of real and imaginary QNMs parts considering $l=2$.

So, analyzing our results, we can verify that the scalar quasinormal modes depend strongly on the field mass, the parameter associated with the nonminimal coupling with gravity, and the loop quantum gravity parameters. Our results show that as the parameter $P$ grows, the real part of the QNMs suffers an initial increase and then starts to decrease, while the magnitude of the imaginary part decreases, considering the fixed-parameter $a_0$. This behavior is also verified in \cite{Santos:2015gja, Cruz:2015bcj, Cruz:2020emz}. This characteristic reveals that the damping of massive scalar perturbations coupling nonminimally with gravity for the self-dual black hole is slower and the oscillations are faster or slower depending on the value of $P$.

Another new phenomenon found here is that for fixed $l$, $n$, and $\xi$, the real and imaginary parts of the quasinormal frequencies increase and decrease, respectively, as the mass of the scalar field, $\mu$, increases. As we know, the $\mu$ has a maximum value, so the QNMs also have a limited value, and due to this fact can be easier to detect by experiments. Also, the introduction of the scalar field mass can help us to understand different aspects of the self-dual black hole, for instance, the aspects of thermodynamics \cite{Ashtekar:2003zx}.

We have verified, that by increasing the value of the nonminimal coupling with gravity, $\xi$, the values of QNMs are modified. Also, for large values of $\xi$, the number of stable modes will significantly decrease. An interesting fact is that for the case when we assume $\mu=0$ and $\xi=1/6$, the obtained results differ from the case $\xi=0$. Therefore, the conformal symmetry can be broken in the presence of LQG corrections. Based on these results, we can conclude that the self-dual black hole has a stable behavior under perturbations provided by a massive scalar field with a nonminimal coupling with gravity.

So, the present study together with \cite{Santos:2015gja, Cruz:2015bcj, Cruz:2020emz} can help us to understand the stability of the self-dual black hole, and also opens a discussion about the interesting issue of conformal symmetry violation in the context of loop quantum black holes. Further analysis considering charged and rotating extensions of SDBH, as well as, other black hole solutions in LQG \cite{Corichi:2015xia, Cortez:2017alh, Olmedo:2017lvt, Yonika:2017qgo, Ashtekar:2018lag, Ashtekar:2018cay, Alesci:2019pbs, Ashtekar:2020ckv}, must improve our understanding of this issue. Also, is very important to consider future analysis applying the extended WKB method to higher orders to the self-dual black hole, well as future investigations must address the issue of quasinormal modes considering the nonminimal coupling with gravity in the context of more recent BHs in LQG.

{\acknowledgements}

We would like to thank CNPq, CAPES and PRONEX/CNPq/FAPESQ-PB (Grant no. 165/2018), for partial financial support. MBC and FAB acknowledge support from CNPq (Grant nos. 150479/2019-0, 312104/2018-9).


\begin{thebibliography}{99}

\bibitem{Einstein:1915ca}
A.~Einstein,
Sitzungsber. Preuss. Akad. Wiss. Berlin (Math. Phys. ) \textbf{1915} (1915), 844-847.

\bibitem{Rovelli:2004tv}
C.~Rovelli, \textit{Quantum gravity}, Cambridge university press, 2004.

\bibitem{Zwiebach:2004tj}
B.~Zwiebach, \textit{A first course in string theory}, Cambridge university press, 2004.

\bibitem{Modesto:2009ve}
L.~Modesto and I.~Premont-Schwarz,
Phys. Rev. D \textbf{80}, 064041 (2009).

\bibitem{LIGOScientific:2018mvr}
B.~P.~Abbott \textit{et al.} [LIGO Scientific and Virgo],
Phys. Rev. X \textbf{9}, no.3, 031040 (2019).

\bibitem{Horowitz:1999jd}
G.~T.~Horowitz and V.~E.~Hubeny,
Phys. Rev. D \textbf{62} (2000), 024027.

\bibitem{Berti:2009kk}
E.~Berti, V.~Cardoso and A.~O.~Starinets,
Class. Quant. Grav. \textbf{26} (2009), 163001.

\bibitem{Dreyer:2002vy}
O.~Dreyer,
Phys. Rev. Lett. \textbf{90} (2003), 081301.

\bibitem{Santos:2015gja}
V.~Santos, R.~V.~Maluf and C.~A.~S.~Almeida,
Phys. Rev. D \textbf{93} (2016) no.8, 084047.

\bibitem{Cruz:2015bcj}
M.~B.~Cruz, C.~A.~S.~Silva and F.~A.~Brito,
Eur. Phys. J. C \textbf{79} (2019) no.2, 157.

\bibitem{Cruz:2020emz}
M.~B.~Cruz, F.~A.~Brito and C.~A.~S.~Silva,
Phys. Rev. D \textbf{102}, no.4, 044063 (2020).

\bibitem{Liu:2020ola}
C.~Liu, T.~Zhu, Q.~Wu, K.~Jusufi, M.~Jamil, M.~Azreg-A\"\i{}nou and A.~Wang,
Phys. Rev. D \textbf{101}, no. 8, 084001 (2020).

\bibitem{Wald:1995yp}
R.~M.~Wald, \textit{Quantum Field Theory in Curved Space-Time and Black Hole Thermodynamics}, University of Chicago press, 1994.

\bibitem{Birrell:1979ip}
N.~D.~Birrell and P.~C.~W.~Davies,
Phys. Rev. D \textbf{22}, 322 (1980).

\bibitem{Konoplya:2002wt}
R.~A.~Konoplya,
Phys. Lett. B \textbf{550}, 117-120 (2002).

\bibitem{Ohashi:2004wr}
A.~Ohashi and M.~a.~Sakagami,
Class. Quant. Grav. \textbf{21}, 3973-3984 (2004).

\bibitem{Konoplya:2018qov}
R.~A.~Konoplya, Z.~Stuchl\'\i{}k and A.~Zhidenko,
Phys. Rev. D \textbf{98}, no.10, 104033 (2018).

\bibitem{Gwak:2019ttv}
B.~Gwak,
Eur. Phys. J. C \textbf{79}, no.12, 1004 (2019).

\bibitem{Yekta:2019por}
D.~Mahdavian Yekta, M.~Karimabadi and S.~A.~Alavi,
[arXiv:1912.12017 [hep-th]].

\bibitem{Spokoiny:1984bd}
B.~L.~Spokoiny,
Phys. Lett. B \textbf{147}, 39-43 (1984).

\bibitem{Barvinsky:2008ia}
A.~O.~Barvinsky, A.~Y.~Kamenshchik and A.~A.~Starobinsky,
JCAP \textbf{11}, 021 (2008).

\bibitem{Setare:2008pc}
M.~R.~Setare and E.~N.~Saridakis,
Phys. Lett. B \textbf{671}, 331-338 (2009).

\bibitem{Uzan:1999ch}
J.~P.~Uzan,
Phys. Rev. D \textbf{59}, 123510 (1999).

\bibitem{Kamenshchik:1995ib}
A.~Y.~Kamenshchik, I.~M.~Khalatnikov and A.~V.~Toporensky,
Phys. Lett. B \textbf{357}, 36-42 (1995).

\bibitem{Donoghue:1994dn}
J.~F.~Donoghue,
Phys. Rev. D \textbf{50}, 3874-3888 (1994).

\bibitem{Allen:1983dg}
B.~Allen,
Nucl. Phys. B \textbf{226}, 228-252 (1983).

\bibitem{Ishikawa:1983kz}
K.~Ishikawa,
Phys. Rev. D \textbf{28}, 2445 (1983).

\bibitem{Maeda:1985bq}
K.~i.~Maeda,
Class. Quant. Grav. \textbf{3}, 233 (1986).

\bibitem{Accetta:1985du}
F.~S.~Accetta, D.~J.~Zoller and M.~S.~Turner,
Phys. Rev. D \textbf{31}, 3046 (1985).

\bibitem{Ashtekar:2003zx}
A.~Ashtekar and A.~Corichi,
Class. Quant. Grav. \textbf{20}, 4473-4484 (2003).

\bibitem{Schutz:1985km}
B.~F.~Schutz and C.~M.~Will,
Astrophys. J. Lett. \textbf{291} (1985), L33-L36.

\bibitem{Iyer:1986np}
S.~Iyer and C.~M.~Will,
Phys. Rev. D \textbf{35} (1987), 3621.

\bibitem{Konoplya:2003ii}
R.~A.~Konoplya,
Phys. Rev. D \textbf{68}, 024018 (2003).

\bibitem{Matyjasek:2017psv}
J.~Matyjasek and M.~Opala,
Phys. Rev. D \textbf{96}, no. 2, 024011 (2017).

\bibitem{Hatsuda:2019eoj}
Y.~Hatsuda,
Phys. Rev. D \textbf{101}, no.2, 024008 (2020).

\bibitem{Konoplya:2019hlu}
R.~A.~Konoplya, A.~Zhidenko and A.~F.~Zinhailo,
Class. Quant. Grav. \textbf{36}, 155002 (2019).

\bibitem{Abreu:1994fd}
J.~P.~Abreu, P.~Crawford and J.~P.~Mimoso, Mannheim:2009qi, 
Class. Quant. Grav. \textbf{11}, 1919-1940 (1994).

\bibitem{Mannheim:2005bfa}
P.~D.~Mannheim,
Prog. Part. Nucl. Phys. \textbf{56}, 340-445 (2006).

\bibitem{Mannheim:2009qi}
P.~D.~Mannheim,
Gen. Rel. Grav. \textbf{43}, 703-750 (2011).

\bibitem{Artymowski:2012is}
M.~Artymowski, A.~Dapor and T.~Pawlowski,
JCAP \textbf{06}, 010 (2013).

\bibitem{Mannheim:2011ds}
P.~D.~Mannheim,
Found. Phys. \textbf{42}, 388-420 (2012).

\bibitem{abramowitz1988handbook}
Abramowitz, Milton and Stegun, Irene A and Romer, Robert H,
American Association of Physics Teachers, 1988.

\bibitem{Berti:2009kk}
E.~Berti, V.~Cardoso and A.~O.~Starinets,
Class. Quant. Grav. \textbf{26}, 163001 (2009).

\bibitem{Kokkotas:1999bd}
K.~D.~Kokkotas and B.~G.~Schmidt,
Living Rev. Rel. \textbf{2}, 2 (1999).

\bibitem{Nollert:1999ji}
H.~P.~Nollert,
Class. Quant. Grav. \textbf{16}, R159-R216 (1999).

\bibitem{Konoplya:2004ip}
R.~A.~Konoplya,
J. Phys. Stud. \textbf{8}, 93-100 (2004).

\bibitem{Simone:1991wn}
L.~E.~Simone and C.~M.~Will,
Class. Quant. Grav. \textbf{9}, 963-978 (1992).

\bibitem{Konoplya:2005hr}
R.~A.~Konoplya,
Phys. Rev. D \textbf{73}, 024009 (2006).

\bibitem{Corichi:2015xia}
A.~Corichi and P.~Singh,
Class. Quant. Grav. \textbf{33}, no.5, 055006 (2016).

\bibitem{Cortez:2017alh}
J.~Cortez, W.~Cuervo, H.~A.~Morales-Técotl and J.~C.~Ruelas,
Phys. Rev. D \textbf{95}, no.6, 064041 (2017).

\bibitem{Olmedo:2017lvt}
J.~Olmedo, S.~Saini and P.~Singh,
Class. Quant. Grav. \textbf{34}, no.22, 225011 (2017).

\bibitem{Yonika:2017qgo}
A.~Yonika, G.~Khanna and P.~Singh,
Class. Quant. Grav. \textbf{35}, no.4, 045007 (2018).

\bibitem{Ashtekar:2018lag}
A.~Ashtekar, J.~Olmedo and P.~Singh,
Phys. Rev. Lett. \textbf{121}, no.24, 241301 (2018).

\bibitem{Ashtekar:2018cay}
A.~Ashtekar, J.~Olmedo and P.~Singh,
Phys. Rev. D \textbf{98}, no.12, 126003 (2018).

\bibitem{Alesci:2019pbs}
E.~Alesci, S.~Bahrami and D.~Pranzetti,
Phys. Lett. B \textbf{797}, 134908 (2019).

\bibitem{Ashtekar:2020ckv}
A.~Ashtekar and J.~Olmedo,
Int. J. Mod. Phys. D \textbf{29}, no.10, 2050076 (2020).

\end{thebibliography}
\end{document}